# All-Optically Controlled Memristor

Lingxiang Hu[1,2], Jing Yang[1], Jingrui Wang[1], Peihong Cheng[1], Leon O. Chua[3] & Fei Zhuge[1,2,4]*

[1]Ningbo Institute of Materials Technology and Engineering, Chinese Academy of Sciences, Ningbo, China

[2]Center of Materials Science and Optoelectronics Engineering, University of Chinese Academy of Sciences, Beijing, China

[3]Department of Electrical Engineering and Computer Sciences, University of California, Berkeley, CA, USA

[4]Center for Excellence in Brain Science and Intelligence Technology, Chinese Academy of Sciences, Shanghai, China

*e-mail: zhugefei@nimte.ac.cn

Memristors have emerged as key candidates for beyond-von-Neumann neuromorphic or in-memory computing[1–8] owing to the feasibility of their ultrahigh-density three-dimensional integration and their ultralow energy consumption. A memristor is generally a two-terminal electronic element with conductance that varies nonlinearly with external electric stimuli and can be remembered when the electric power is turned off[9]. As an alternative, light can be used to tune the memconductance and endow a memristor with a combination of the advantages of both photonics and electronics[10,11]. Both increases[12–14] and decreases[15,16] in optically induced memconductance have been realized in different memristors; however, the reversible tuning of memconductance with light in the same device remains a considerable challenge that severely restricts the development of optoelectronic memristors. Here we describe an all-optically controlled (AOC) analog memristor with memconductance that is reversibly tunable over a continuous range by varying only the wavelength of the controlling light. Our memristor is based on the relatively mature semiconductor material InGaZnO (IGZO) and a memconductance tuning mechanism of light-induced electron trapping and detrapping. We demonstrate that spike-timing-dependent plasticity (STDP) learning can be realized in our device, indicating its potential applications in AOC spiking neural networks (SNNs) for highly efficient optoelectronic neuromorphic computing.

A memristor generally consists of an insulating or semiconducting thin film sandwiched between two electrodes. Memconductance can be reversibly tuned with electric stimuli via various mechanisms, such as field-driven ion migration[17–24] or electron trapping[25] and current-induced phase change[26], depending on the film and electrode materials used. Light has also been found to be an effective approach for the modulation of memconductance[12–16,27,28], hence enabling the realization of optoelectronic memristors. For an ideal optoelectronic memristor[29,30], its memconductance should be all-optically tunable and electrically readable. However, to date, optoelectronic memristors have been realized only through a combination of optical and electrical stimuli[13–16,27]. Such a controlling scheme will increase the operation complexity of memristors and thus hinder their usage in optoelectronic applications. For example, for memristive devices based on electron trapping[12,13] or proton intercalation[14], a light-induced persistent photocurrent[12,13] or photocatalytic[14] effect was used to increase the memconductance; however, a decrease in memconductance could be achieved only via electric stimuli[13,14] or with strong dependence on an external electric field[27]. In the case of memristors based on conducting nanofilaments (belonging to the class of ion-migration-driven memristors), light illumination can cause filament breakage, resulting in decreased memconductance[15,16]; however, to increase the memconductance via filament formation or rejuvenation, electric excitation is still necessary[15,16]. In addition, a photomechanical switching effect can be used to modulate memconductance[28], although this causes severe expansion or contraction of the device. To the best of our knowledge, the reversible tuning of memconductance by applying only optical excitation while simultaneously leaving the device structure unchanged has not yet been realized.

The main goal of this study was to realize an AOC memristor. To achieve this goal, we used a wide-band-gap amorphous oxide material, IGZO, as the active layer. IGZO is widely used as a key thin-film transistor material for high-performance display production. The device is based on an oxygen-deficient IGZO ($O_D$-IGZO)/oxygen-rich IGZO ($O_R$-IGZO) homojunction (see the inset of Fig. 1a, Methods and Extended Data Fig. 1). Such a bilayered structure is crucial for achieving an AOC memristor, as will be discussed in detail later. This device demonstrates typical memristive behavior[31] when measured in the dark (Fig. 1a). Positive voltage sweeping converts the device from a low memconductance state (LMS) to a high memconductance state (HMS), referred to as the SET operation; the reverse process of memconductance switching from the HMS to the LMS under negative voltage sweeping is called the RESET operation. The LMS and HMS exhibit nonvolatility (see Extended Data Fig. 2a). The memconductance can also be reversibly modulated by alternately applying positive and negative voltage pulses (see Extended Data Fig. 2b).

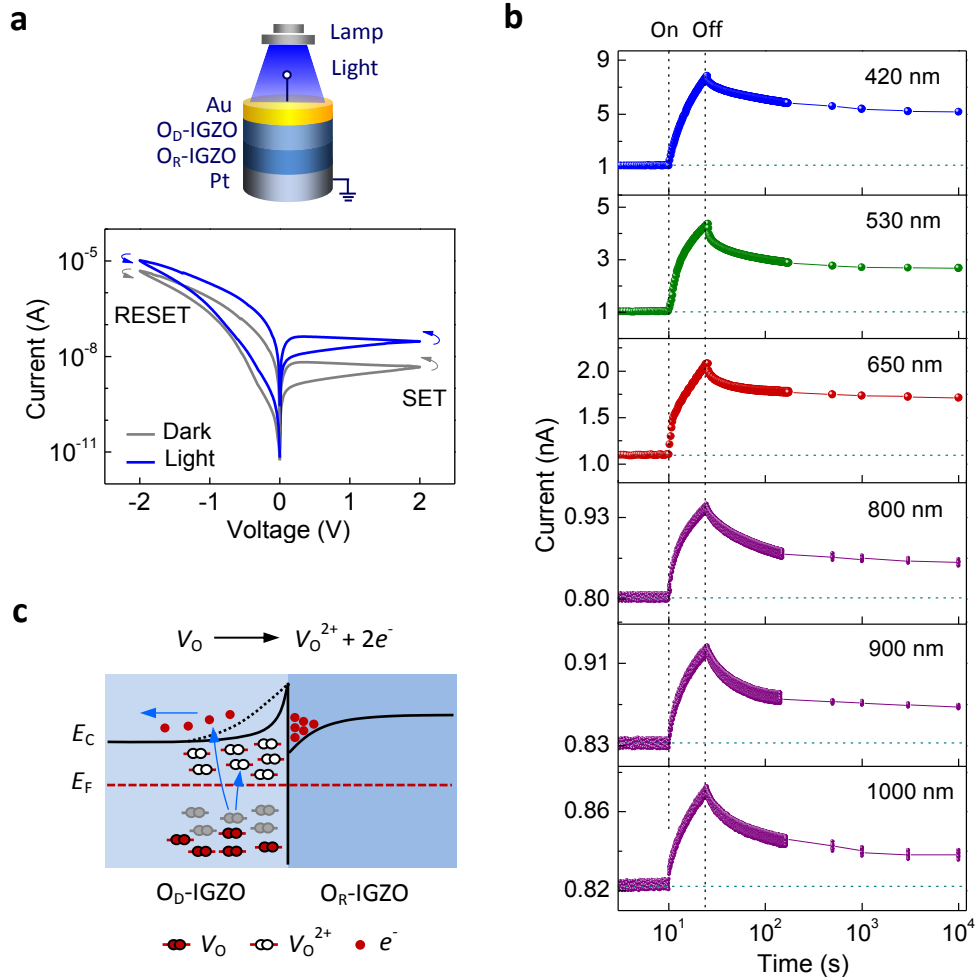

**Fig. 1 | Memristive switching and optical SET behavior. a,** Current–voltage characteristics before and after light irradiation (wavelength ($\lambda$) = 420 nm, duration ($D$) = 15 s and power density ($P$) = 20 $\mu$W/cm$^2$). The inset shows a schematic diagram of the measurement configuration. **b,** Optical SET behavior upon irradiation with light of various wavelengths ($D$ = 15 s and $P$ = 20 $\mu$W/cm$^2$). The horizontal dashed lines indicate the initial LMS. The vertical dashed lines indicate the times at which the light is switched on and off. The current values were measured at 10 mV. **c,** Equilibrium energy band diagrams of the O$_D$-IGZO/O$_R$-IGZO interface after the optical SET operation. The $V_O$ ionization (blue arrows) reaction is also schematically illustrated. $E_F$ and $E_C$ denote the Fermi energy and conduction band minimum, respectively. The black dashed lines indicate the positions of $E_C$ before the SET operation. Note that the reaction actually occurs under nonequilibrium conditions.

To determine the mechanism of memristive switching, we analyzed the band structures of O$_D$-IGZO and O$_R$-IGZO (for details, see Methods and Extended Data Fig. 3). The analysis indicates that the electrons in $O_D$-IGZO tend to diffuse into $O_R$-IGZO, resulting in the formation of a built-in electric field at the $O_D$-IGZO/$O_R$-IGZO interface and, thus, a potential barrier on the $O_D$-IGZO side and a potential well on the $O_R$-IGZO side. The width of this interfacial barrier, which depends on the density of ionized oxygen vacancies ($V_O^{2+}$s) and determines the tunneling current[32,33], plays a key role in memristive switching (for details, see Methods and Extended Data Fig. 4). Specifically, the SET behavior originates from electron detrapping at neutral oxygen vacancies ($V_O$s) located in the interfacial barrier region; that is, ionization of the $V_O$s causes an increase in the density of $V_O^{2+}$s, thus leading to a decrease in the width of the barrier, which facilitates electron tunneling across the junction (see Extended Data Fig. 4e). In contrast, for the RESET behavior, electron trapping at $V_O^{2+}$s, *i.e.*, the neutralization of $V_O^{2+}$s, gives rise to an increase in the width of the barrier, thus resulting in a lowered tunneling current (see Extended Data Fig. 4f).

Having demonstrated memristive behavior under electrical stimulation, we investigated the performance under light exposure. The Au/$O_D$-IGZO/$O_R$-IGZO structure shows a transmittance of > 55% for light wavelengths from 400 to 1000 nm (see the inset in Extended Data Fig. 5a). Therefore, the device can be irradiated with visible (*e.g.*, 420–650 nm) and near-infrared (*e.g.*, 800–1000 nm) light for memconductance modulation. When the device was measured soon after blue light (420 nm) exposure, we observed memristive behavior similar to that observed in the dark, but with a significant increase in current (Fig. 1a). To better understand the influence of irradiation on the memconductance, the effects of irradiating the device using light of various wavelengths (420, 530, 650, 800, 900 and 1000 nm) were investigated (Fig. 1b). The current gradually increases under these sub-band-gap illumination conditions. After irradiation, the device exhibits a strong persistent photocurrent. This shows that both visible and near-infrared light can convert the device to a nonvolatile HMS. The light-induced HMS can be reset to the LMS by means of electric stimuli. Thus, the memconductance can be reversibly modulated through a combination of visible (or near-infrared) light and negative voltage pulses (see Extended Data Fig. 5).

Single-layered $O_D$-IGZO or $O_R$-IGZO devices exhibit markedly different behavior, which provides insight into the mechanism of optical memconductance. No photocurrent can be observed upon visible light irradiation of the $O_D$-IGZO device (see Extended Data Fig. 6a). For the $O_R$-IGZO device, a volatile photocurrent is generated upon visible light illumination, but no photocurrent appears upon near-infrared light irradiation (see Extended Data Fig. 6b). We therefore deduce that for the bilayered $O_D$-IGZO/$O_R$-IGZO device, the interfacial barrier region plays a key role in the optical SET operation (Fig. 1c). Specifically, light-induced transformation from the $V_{OS}$ located in the interfacial barrier region into $V_O^{2+}$s causes a decrease in the width of the barrier, thus giving rise to an increased memconductance. The optoelectronic response to near-infrared light indicates the existence of $V_{OS}$ with energy levels as shallow as 1.24 eV (1000 nm wavelength) below the conduction band of $O_D$-IGZO. Such shallow energy levels may arise from a high $V_O$ density, which can broaden the energy level distribution of the $V_{OS}$[34]. The unique long-wavelength response capability of the $O_D$-IGZO/$O_R$-IGZO device enables the realization of an AOC memristor.

The key to achieving an AOC memristor is to realize an optical RESET operation. To achieve this goal, the device was first exposed to blue light (420 nm) for 15 s to set it to an HMS. Then, it was irradiated with light of various wavelengths from 530 to 900 nm at 10 minutes after the initial blue light exposure (Fig. 2a). Green light (530 nm) converted the device to even higher HMSs, whereas red light (650 nm) resulted in a weak decrease in the photocurrent after an initial increase. Upon near-infrared light irradiation (800 and 900 nm), a strong decrease in the current, *i.e.*, RESET behavior, was observed following a relatively weak increase; near-infrared light at 800 nm was more efficient than 900 nm light for reducing the device current. Therefore, we selected 800 nm for subsequent RESET operations in pulse mode (Fig. 2b). As demonstrated in Fig. 2b, the RESET efficiency increased with the light power density.

We also found that the wavelength range for the RESET operation strongly depends on the light used for the initial SET operation. After ultraviolet light illumination (350 nm), the device can be reset to an LMS by both red and near-infrared light (see Extended Data Fig. 7a). However, a device irradiated with green or red light cannot be restored to the LMS even by near-infrared light (see Extended Data Fig. 7b, c). Moreover, the ultraviolet-light-illuminated device (see Extended Data Fig. 7a) exhibits a current reduction of a larger magnitude upon near-infrared light irradiation than the blue-light-exposed device in Fig. 2a. These findings suggest that setting the device with short-wavelength light facilitates the subsequent RESET operation.

Because the optical SET mechanism depends on barrier narrowing (Fig. 1c), we expected that widening of the barrier region due to a reduced $V_O^{2+}$ density would result in an optical RESET. Upon near-infrared light irradiation, electrons in the potential well formed by band bending of O$_R$-IGZO are apt to tunnel through[35] or jump over[36] the barrier and enter the O$_D$-IGZO conduction band (Fig. 2c). Some electrons are captured by $V_O^{2+}$s, which then transform into $V_O$s. This mechanism can explain the dependence of the RESET efficiency on the light power density in Fig. 2b; that is, light with a higher power density excites more electrons into the conduction band of O$_D$-IGZO, leading to a larger probability of $V_O^{2+}$ neutralization and thus widening the interfacial barrier.

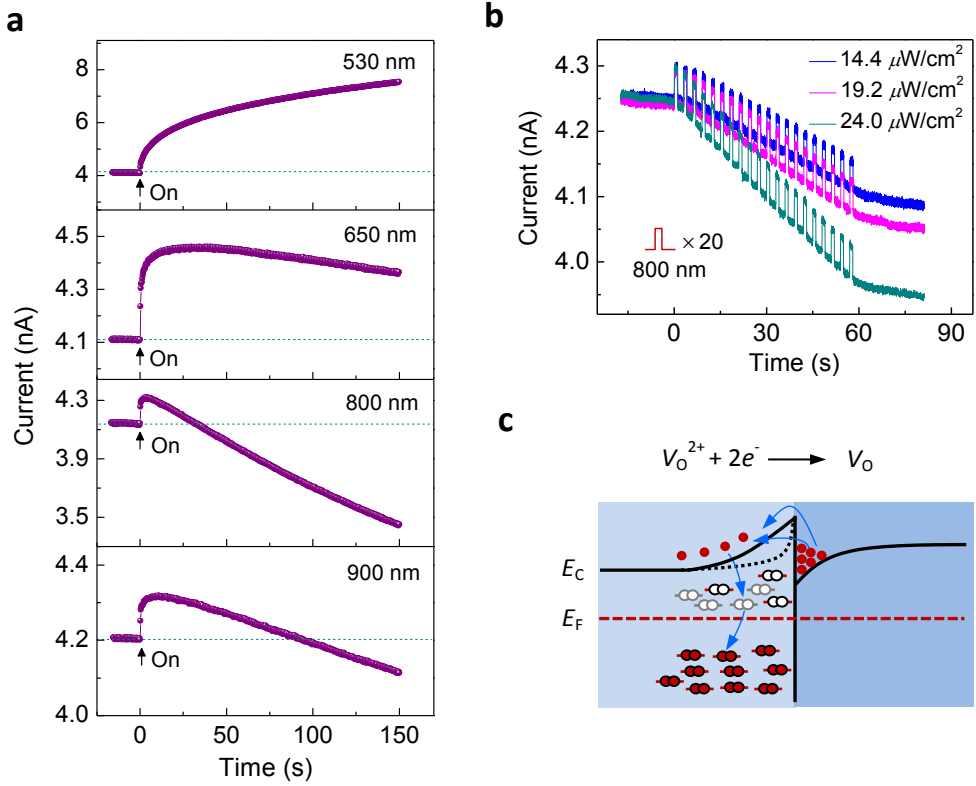

**Fig. 2 | Optical RESET behavior. a,** Photocurrent responses to irradiation with light of various wavelengths ($P = 20 \mu$W/cm$^2$). The device was first set to an HMS by blue light irradiation ($D = 15$ s and $P = 20 \mu$W/cm$^2$). The horizontal dashed lines indicate the initial HMS. The vertical dashed lines indicate the times at which the light is switched on. **b,** Optical RESET behavior upon exposure to light pulses of various power densities ($D = 1$ s and pulse interval ($I$) $= 2$ s). Before applying pulses, the device underwent the same SET operation as in **a**. In **a** and **b**, the current values were measured at 10 mV. **c,** Equilibrium energy band diagram after the optical RESET operation. The electron tunneling and jumping and subsequent $V_O^{2+}$ neutralization processes that occur upon irradiation are also schematically illustrated (blue arrows). The black dashed lines indicate the positions of $E_C$ before the RESET operation. Note that the reaction actually occurs under nonequilibrium conditions.

We further propose that upon illumination, these two opposite reactions, namely, the ionization of $V_O$s and the neutralization of $V_O^{2+}$s, occur simultaneously. The photocurrent depends on the dynamic equilibrium between these two reactions. In the ionization-dominated (or neutralization-dominated) case, the device shows an increase (or decrease) in photocurrent. For a device in the LMS, the ionization of $V_O$s is dominant, thus resulting in an increase in memconductance upon light irradiation (Fig. 1b). However, when the device is initially exposed to light of a relatively short wavelength, only a small number of $V_O$s can be ionized by subsequent irradiation at a longer wavelength owing to the low density of $V_O$s with relatively shallow energy levels. In this case, the neutralization of $V_O^{2+}$s may dominate, enabling a decrease in the memconductance (Fig. 2a, b and Extended Data Fig. 7a, b). The increase (or initial increase) in the memconductance observed in Fig. 2a, b and Extended Data Fig. 7 most likely originates from the ionization of $V_O$s regenerated from the spontaneous neutralization of $V_O^{2+}$s by electron tunneling through the interfacial barrier. This spontaneous transformation from $V_O^{2+}$s to $V_O$s gives rise to the persistent photocurrent decay that was seen in Fig. 1b.

We implemented our bilayered $O_D$-IGZO/$O_R$-IGZO memristor using blue and near-infrared light pulses for the SET and RESET processes, respectively, as schematically illustrated in Fig. 3a. The top panel in Fig. 3b shows a continuous increase and decrease in the memconductance under a series of successive light pulses. Our AOC memristor exhibits good operation endurance (Fig. 3b, bottom panel). To verify the nonvolatility of the light-induced memconductance states, we present retention measurements of ten states obtained after both the SET and RESET operations (Fig. 3c). The memconductance exhibits an initial slow decay, which is an intrinsic feature of a persistent photocurrent, and then remains stable, with all states being clearly distinguishable even after $10^4$ s. As mentioned above, the persistent photocurrent decay is related to the tunneling of electrons through the interfacial barrier, which is why the decay was weaker for lower memconductance. More specifically, a lower memconductance implies a wider barrier, resulting in fewer electrons tunneling through it. Subsequently, fewer $V_O^{2+}$s transform into $V_O$s, thus leading to less barrier widening.

Our AOC memristor is structure-changeable. For example, a transparent conducting oxide (Sn-doped $In_2O_3$ (ITO)) can also be used as the top electrode. The ITO/$O_D$-IGZO/$O_R$-IGZO/Pt demonstrates electrical and optoelectronic behavior similar to that of Au/$O_D$-IGZO/$O_R$-IGZO/Pt (see Extended Data Fig. 8).

The multiple reversibly tunable memconductance states of our AOC memristor (Fig. 3b)

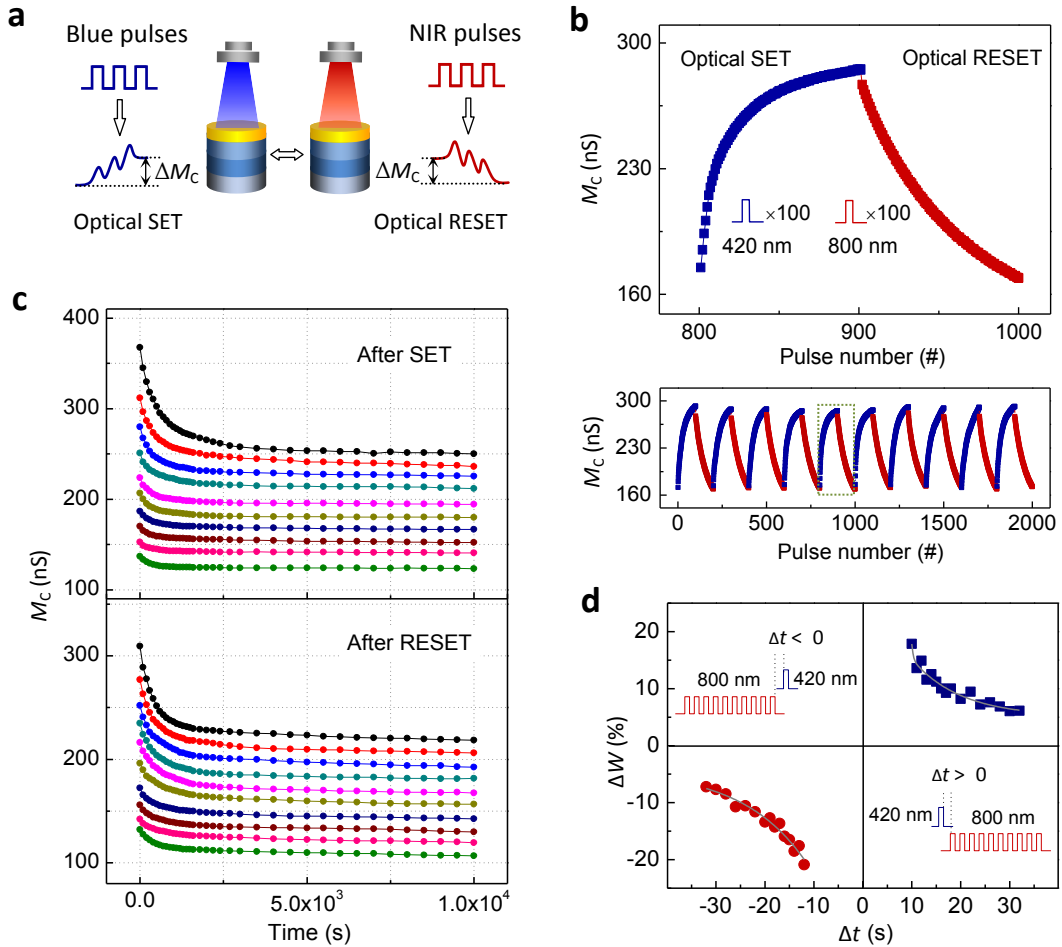

**Fig. 3 | AOC memristor and its application in neuromorphic computing. a,** Schematic of the realization of an AOC memristor. $M_C$ and NIR denote the memconductance and near-infrared, respectively. **b,** Top panel: Reversible regulation of the memconductance by means of 100 blue light pulses ($D$ = 1 s, $I$ = 1 s and $P$ = 20 $\mu$W/cm$^2$) and 100 NIR light pulses ($D$ = 1 s, $I$ = 1 s and $P$ = 24 $\mu$W/cm$^2$). Bottom panel: Ten successive memconductance increase–decrease cycles. The top panel shows an enlarged view of the fifth cycle, as marked by the green rectangle in the bottom panel. The memconductance values were measured 1 s after each light pulse. **c,** Retention characteristics of ten memconductance states after both optical SET (top panel) and optical RESET (bottom panel) operations. **d,** Emulation of synaptic STDP in the AOC memristor. The insets illustrate the light pulse schemes. In **b**, **c** and **d**, the memconductance values were measured at 10 mV.

make it an excellent synaptic emulator. Specifically, we found that our memristor can mimic STDP — an important learning rule in the brain that requires variation in the synaptic weight ($W$) to be a strong function of the pre- and postneuron spike timing ($\Delta t$)[37] (Fig. 3d). A single blue light pulse and a train of ten near-infrared light pulses serve as pre- and postsynaptic spikes, respectively (Fig. 3d, inset). The memconductance is denoted as $W$. We see that synaptic potentiation ($\Delta W > 0$) occurs when the presynaptic spike arrives before the postsynaptic spike ($\Delta t > 0$); by contrast, synaptic depression ($\Delta W < 0$) occurs when the postsynaptic spike arrives first ($\Delta t < 0$). In addition, $|\Delta W|$ increases with decreasing $|\Delta t|$. These observations are consistent with the typical STDP characteristics of biological synapses[37]. Given that STDP is the basic learning rule of SNNs that reflect the information processing style in the human brain[38], our AOC memristor is promising for applications in optoelectronic SNNs. The mechanism of STDP emulation is explained in detail in the Methods and Extended Data Fig. 9.

In conclusion, the realization of AOC memristors is an important milestone towards optoelectronic applications of memristors, such as neuromorphic computing. Optoelectronic computing using our AOC memristor is more practically feasible than purely optical computing[10,39] owing to the simple structure and easy fabrication of this device. Future research might explore convenient methods for controllably introducing light into individual cells in high-density memristor crossbar arrays, as well as neuronal functions (*e.g.*, integrate and fire) using our AOC memristor, to enable optoelectronic SNNs. It is worth noting that, strictly speaking, our memristor is semi-nonvolatile rather than fully nonvolatile, given the spontaneous decay of photocurrents (Fig. 3c), which is an intrinsic feature of optoelectronic devices[12–14,29,33,40]. Such semi-nonvolatility is consistent with the evolution tendency of memory in the human brain[41]; however, it may affect the computing accuracy of artificial neural networks[42], which might also need to be taken into consideration in future research.

## Methods

### Material growth

Amorphous IGZO thin films with a diameter of 100 $\mu$m were deposited on Pt/Ti/SiO$_2$/Si, SiO$_2$/Si and quartz substrates at room temperature (RT) via RF magnetron sputtering of an InGaZnO$_4$ (In$_2$O$_3$:Ga$_2$O$_3$:ZnO = 1:1:2, molar ratio) ceramic target of 99.99% purity with *in situ* metal shadow masks. The sputtering power and pressure were 60 W and 0.5 Pa, respectively. O$_D$-IGZO was sputtered in pure Ar gas, and O$_R$-IGZO was deposited in a mixed Ar and O$_2$ atmosphere with a partial pressure ratio of 1:1. The thickness of the O$_D$-IGZO and O$_R$-IGZO films was ~30 nm. For the fabrication of O$_D$-IGZO/O$_R$-IGZO homojunctions, O$_R$-IGZO was sputtered first, followed by deposition of O$_D$-IGZO.

### Device fabrication

Ten-nanometer-thick Au top electrodes with a diameter of 100 $\mu$m were deposited onto O$_D$-IGZO, O$_R$-IGZO and O$_D$-IGZO/O$_R$-IGZO films at RT via electron-beam evaporation with *in situ* metal shadow masks. ITO electrodes with a thickness of 30 nm and a diameter of 100 $\mu$m were deposited at RT via RF magnetron sputtering of an ITO (In$_2$O$_3$:SnO$_2$ = 5:1, molar ratio) ceramic target of 99.99% purity with *in situ* metal shadow masks. Ar gas was used as the sputtering atmosphere. The sputtering power and pressure were 40 W and 0.5 Pa, respectively.

## Electrical and optoelectronic characterization

Electrical and optoelectronic measurements were performed at RT in air using a Keithley 4200 semiconductor parameter analyzer equipped with a monochromatic light source (Omni-λ 3007). The bias voltage was applied to the top electrode (Au or ITO) with the bottom electrode (Pt) grounded, and the light entered into the device through the top electrode (Au or ITO) (Fig. 1a).

We found that the as-fabricated bilayered $O_D$-IGZO/$O_R$-IGZO device was not in an LMS but in an HMS with a memconductance of approximately $10^2$–$10^3$ nS (at −10 mV). Given that the device was highly sensitive to visible light and that exposure to environmental light was inevitable, such an HMS should be induced by environmental light exposure. To restore the device to the initial LMS before electrical or optoelectronic measurements, an electrical or optical RESET operation was first performed. After the RESET operation, the device presented a low memconductance of approximately $10^{-1}$–$10^2$ nS (at −10 mV) depending on the RESET parameters used.

## Material characterization

The film thickness was measured via variable angle spectroscopic ellipsometry (M-2000 DI, J. A. Woollam Co., Inc.). The carrier density and resistivity were determined with a Hall effect measurement system (HP-5500C, Nanometrics) using the van der Pauw method. Absorption and transmittance spectra were collected using both a UV−visible−IR spectrophotometer (Lambda 950, PerkinElmer) and spectroscopic ellipsometry. All of the above measurements were performed at RT in air. Cross-sectional microstructural observations were performed by means of high-resolution transmission electron microscopy (HRTEM, Talos F200X, Thermo Fisher). Cross-sectional specimens were fabricated by means of focused ion beam etching. Elemental and energy-band structure analyses were conducted using X-ray photoelectron spectroscopy (XPS, Kratos Axis Ultra DLD). The work function was determined via ultraviolet photoelectron spectroscopy (UPS). $O_D$-IGZO and $O_R$-IGZO films with a thickness of 100 nm were used for the XPS and UPS measurements. To eliminate the influence of surface contamination (*e.g.*, moisture from the atmosphere) on the measurements, an ~30-nm-thick surface layer was etched with Ar plasma before the optoelectronic signals were collected.

The HRTEM images and corresponding fast Fourier transform (FFT) images revealed the amorphous structures of $O_D$-IGZO and $O_R$-IGZO (Extended Data Fig. 1). From the XPS measurements, the molar ratios of In:Ga:Zn:O for the $O_D$-IGZO and $O_R$-IGZO were estimated to be 1.2 : 2.2 : 1 : 4.5 and 1.2 : 2.1 : 1 : 4.6, respectively.

As determined by Hall effect measurements, $O_D$-IGZO showed an electron concentration of $10^{19}$ cm$^{-3}$ and a resistivity of $10^{-2}$ Ωcm. $O_R$-IGZO had a much higher resistivity, which exceeded the measurement limit (~$10^5$ Ωcm) of the Hall measurement system. Given that oxygen vacancies are the main n-type dopants in IGZO[43], the low resistivity of the $O_D$-IGZO indicates a high density of oxygen vacancies. The lower conductivity of $O_R$-IGZO indicates that it contains much fewer oxygen vacancies than $O_D$-IGZO. The oxygen vacancies in IGZO have a wide distribution of energy levels in the band gap[44–46].

**Band structure analysis**

The optical band gaps ($E_g$) of $O_D$-IGZO and $O_R$-IGZO were determined to be 3.7 and 3.6 eV, respectively, via a modified Kubelka-Munk function (Extended Data Fig. 3a). The $E_g$ values were obtained from the abscissa intercepts of the straight lines fitted to the linear portions of the plotted data points in the high energy region.

The differences between the Fermi energy ($E_F$) and the valence band maximum ($E_V$) for $O_D$-IGZO and $O_R$-IGZO were determined to be 3.1 and 2.5 eV, respectively, by measuring the valence band XPS spectra (Extended Data Fig. 3b). The values were obtained from the positions of the intersections of the straight lines fitted to the leading edges of the spectra and the straight lines fitted to the flat energy distribution portions in the low energy region.

The work functions ($\Phi$) of $O_D$-IGZO and $O_R$-IGZO were determined to be 4.2 and 4.9 eV, respectively, by measuring the valence band UPS spectra (He I, 21.22 eV) (Extended Data Fig. 3c). The position of the secondary electron cut-off ($E_{cut\text{-}off}$) was determined from the midpoint of the cut-off edge. $\Phi$ could then be calculated as $\Phi = h\nu - E_{cut\text{-}off}$, where $h$ is the Planck constant, $\nu$ is the frequency of the light and $h\nu$ is the photon energy (21.22 eV). The differences between $E_F$ and $E_V$ for $O_D$-IGZO and $O_R$-IGZO were determined to be 3.2 and 2.6 eV, respectively, which are consistent with the values obtained from the XPS spectra (Extended Data Fig. 3b).

We can then plot the energy band diagrams of $O_D$-IGZO and $O_R$-IGZO before contact, as shown in Extended Data Fig. 3d. The differences between the conduction band minimum ($E_C$) and $E_F$ for $O_D$-IGZO and $O_R$-IGZO were calculated to be 0.6 and 1.1 eV, respectively, according to $E_g - (E_F - E_V)$. Subsequently, the differences between the vacuum energy ($E_{VAC}$) and $E_C$ for $O_D$-IGZO and $O_R$-IGZO were determined to be 3.6 and 3.8 eV, respectively, according to $\Phi - [E_g - (E_F - E_V)]$.

After contact, the Fermi levels of $O_D$-IGZO and $O_R$-IGZO tend to equilibrate via electron transfer from $O_D$-IGZO to $O_R$-IGZO (Extended Data Fig. 3e), resulting in the formation of positive space charges (ions) on the $O_D$-IGZO side and negative space charges (electrons) on the $O_R$-IGZO side and thus a built-in electric field at the $O_D$-IGZO/$O_R$-IGZO interface (Extended Data Fig. 3f). As a consequence, the energy band is bent upwards on the $O_D$-IGZO side and downwards on the $O_R$-IGZO side; that is, a potential barrier is formed on the $O_D$-IGZO side, and a potential well is formed on the $O_R$-IGZO side (Extended Data Fig. 3e).

**Analysis of the memristive switching mechanism**

Given the pronounced difference in current–voltage ($I$–$V$) characteristics between a bilayered $O_D$-IGZO/$O_R$-IGZO device (Fig. 1a) and a single-layered $O_D$-IGZO or $O_R$-IGZO device (Extended Data Fig. 4a, b), we deduce that the $O_D$-IGZO/$O_R$-IGZO interfacial region plays a key role in memristive switching. Moreover, the strong dependence of the memconductance on the device size (*i.e.*, the diameter of the top electrode or IGZO thin film) for both HMSs and LMSs indicates radially homogeneous switching behavior[47] (Extended Data Fig. 4c). We therefore propose that the memristive switching of a bilayered $O_D$-IGZO/$O_R$-IGZO device originates from electron trapping (or detrapping) at ionized (or neural) oxygen vacancies ($V_O^{2+}$s or $V_O$s) located in the $O_D$-IGZO/$O_R$-IGZO interfacial barrier region (Extended Data Fig. 4e, f). This proposal is consistent with the memconductance tuning mechanism upon light irradiation.

More specifically, we propose that the memconductance of the device mainly depends on the width of the $O_D$-IGZO/$O_R$-IGZO interfacial barrier, which determines the tunneling current[32,33]. The barrier width is determined by the density of positive ion space charges (*i.e.*, ionized oxygen vacancies): a higher density of ionized oxygen vacancies results in a narrower barrier width[32]. As mentioned above, there are abundant $V_O$s with a wide distribution of energy levels in $O_D$-IGZO. The $V_O$s in relatively shallow energy states tend to be doubly ionized[48,49], transforming into $V_O^{2+}$s with shallow energy levels (Extended Data Fig. 4d). When a positive bias is applied to the device, some of the $V_O$s in relatively deep energy states ionize into $V_O^{2+}$s[50] (Extended Data Fig. 4e). These additional $V_O^{2+}$s generated in the barrier region cause the barrier to narrow, which facilitates electron tunneling across the junction. Thus, the device is set to an HMS. The HMS is nonvolatile because free electrons produced in the barrier region are swiftly pulled into $O_D$-IGZO by either the built-in or the external electric field and therefore cannot recombine with $V_O^{2+}$s. The existence of an energy barrier hindering $V_O^{2+}$ neutralization also contributes to the nonvolatility[48]. When a negative voltage is applied, the interfacial barrier is forward biased, and many electrons are injected into the barrier region. Some electrons will be captured by $V_O^{2+}$s generated during the electrical SET operation, which then transform back into $V_O$s (Extended Data Fig. 4f). As a result, the barrier becomes wide again and the device is reset to the LMS.

The current decay of the HMS in Extended Data Fig. 2a can therefore be explained by the spontaneous neutralization of ionized oxygen vacancies: some electrons tunneling through the interfacial barrier from $O_R$-IGZO to $O_D$-IGZO are captured by ionized oxygen vacancies, which then transform into neutral oxygen vacancies; a decrease in the number of ionized oxygen vacancies results in widening of the interfacial barrier and, thus, a reduction in current.

**Analysis of the mechanism for emulating STDP**

As shown in Extended Data Fig. 9, $M_{C0}$ is the initial memconductance value, and $M_{C1}$ and $M_{C2}$ are the updated values that are measured 60 s after the spikes. The synaptic weight variation was calculated using the following formula: $\Delta W_i = [(M_{Ci} - M_{C0})/M_{C0}] \times 100\%$, where $i = 1$ and 2.

For the emulation of synaptic potentiation, the device was initially set to a relatively low memconductance of $M_{C0} \approx 100$ nS. Extended Data Fig. 9a presents two $W$ modulation events with $\Delta t_2 > \Delta t_1 > 0$. First, the blue light pulse induces a positive persistent photoconductance, that is, a light-induced memconductance that is higher than $M_{C0}$. The subsequent near-infrared light pulses induce a negative persistent photoconductance, that is, a light-induced memconductance that is lower than that before near-infrared irradiation. We can see from Extended Data Fig. 9a that the updated memconductance ($M_{C1}$ or $M_{C2}$) is higher than $M_{C0}$, indicating the realization of synaptic potentiation. Given that $\Delta t_1 < \Delta t_2$, the $M_C$ values before and after near-infrared light irradiation for $\Delta t_1$ are higher than those for $\Delta t_2$. It follows that $M_{C1} > M_{C2}$, and thus, $\Delta W_1 > \Delta W_2$.

To mimic synaptic depression, the device was initially set to a relatively high memconductance of $M_{C0} \approx 250$ nS. Extended Data Fig. 9b shows two $W$ modulation events with $\Delta t_2 < \Delta t_1 < 0$. First, the near-infrared light pulses induce a negative persistent photoconductance. The subsequent blue light pulse induces a positive persistent photoconductance. The updated memconductance ($M_{C1}$ or $M_{C2}$) is lower than $M_{C0}$, indicating the realization of synaptic depression. Given that $|\Delta t_1| < |\Delta t_2|$, the memconductance before blue light irradiation for $\Delta t_1$ is higher than that for $\Delta t_2$. However, because of the relatively large decay amplitude of the memconductance after blue light irradiation, the subsequent memconductance for $\Delta t_1$ is lower than that for $\Delta t_2$. Therefore, $M_{C1} < M_{C2}$ and $|\Delta W_1| > |\Delta W_2|$.

## Data availability

The data that support the findings of this study are available from the corresponding authors upon reasonable request.

**Acknowledgements** This work is supported in part by the National Natural Science Foundation of China (Nos. 61674156 and 61874125), the Strategic Priority Research Program of Chinese Academy of Sciences (No. XDB32050204) and the Zhejiang Provincial Natural Science Foundation of China (No. LD19E020001).

**Author contributions** L.H. and F.Z. conceived and designed the experiments. L.H., J.Y. and P.C. conducted the experiments. L.H., J.W., L.O.C. and F.Z. prepared the manuscript. F.Z. supervised the project. All authors discussed the results and commented on the manuscript.

**Competing interests** The authors declare no competing interests.

**Additional information**

**Correspondence and requests for materials** should be addressed to F.Z.

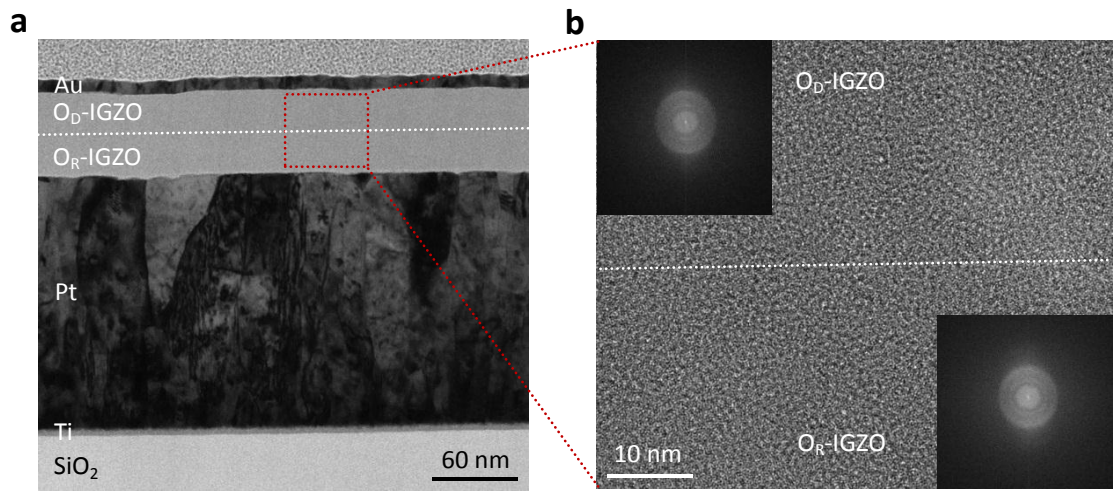

**Extended Data Fig. 1 | Structure and microstructure of an $O_D$-IGZO/$O_R$-IGZO thin film. a,** TEM image of a Au/$O_D$-IGZO/$O_R$-IGZO/Pt device deposited on Pt/Ti/SiO$_2$/Si. **b,** HRTEM image of the area marked by the red rectangle in **a**. The insets present the corresponding FFT images. In **a** and **b**, the white dotted lines indicate the boundary between $O_D$-IGZO and $O_R$-IGZO.

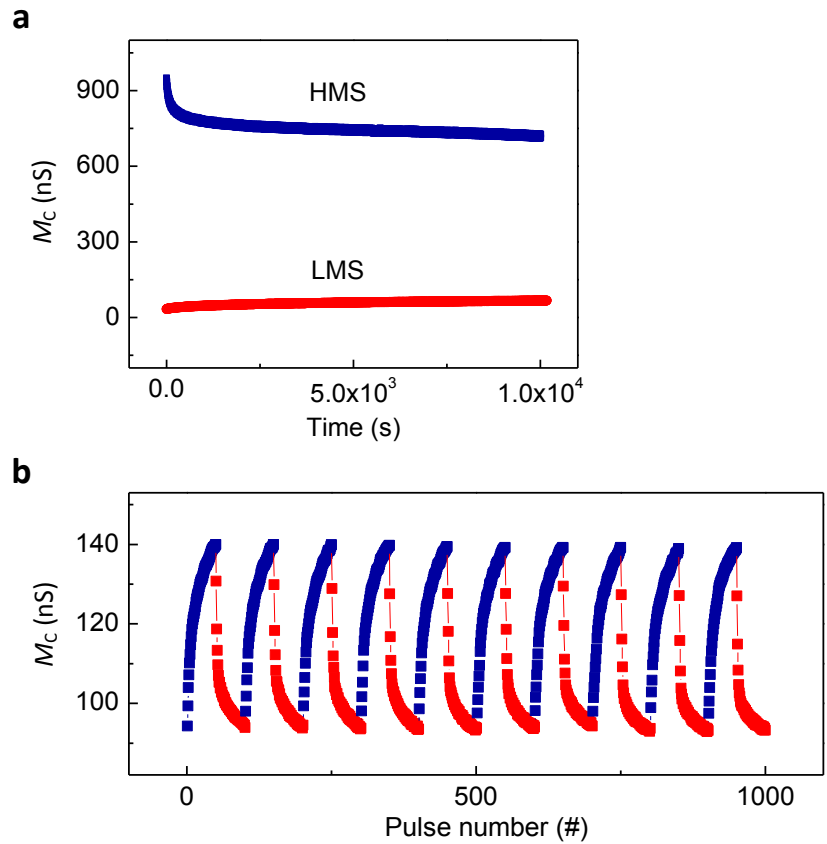

**Extended Data Fig. 2 | Memristive switching performance of Au/O$_D$-IGZO/O$_R$-IGZO/Pt. a,** Retention of the LMS and HMS. **b,** Reversible regulation of the memconductance by means of 50 positive voltage pulses (SET, blue) and 50 negative voltage pulses (RESET, red) with an amplitude of 2 V, a duration of 25 ms and an interval of 250 ms. In **a** and **b**, the memconductance values were obtained at 10 mV.

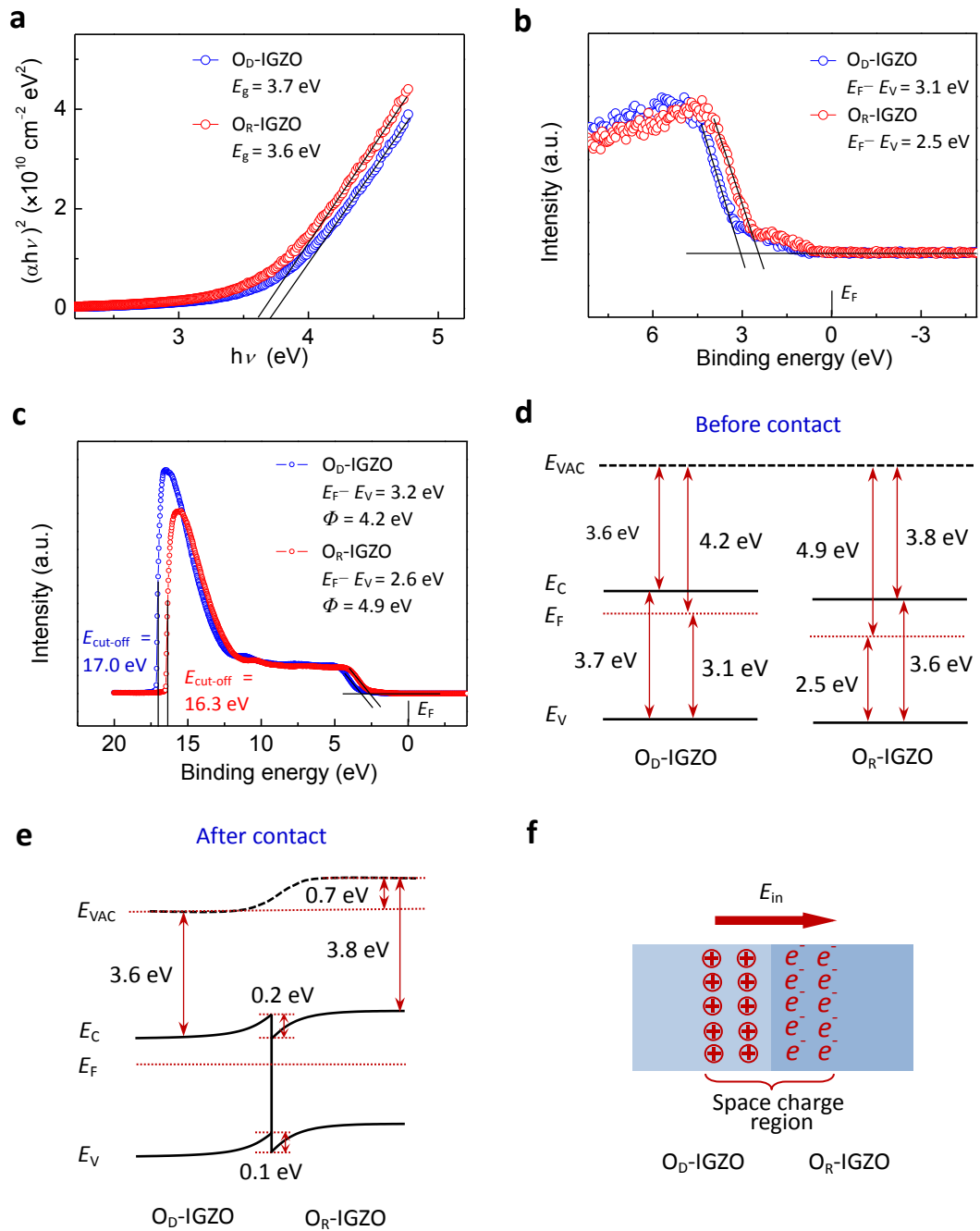

**Extended Data Fig. 3 | Band structure analysis of $O_D$-IGZO and $O_R$-IGZO. a,** $(\alpha h\nu)^2$ versus $h\nu$ plots for $O_D$-IGZO and $O_R$-IGZO, where $\alpha$ is the light absorption coefficient, $h$ is the Planck constant and $\nu$ is the frequency of the light. The $\alpha$ values were measured by means of a UV–visible–IR spectrophotometer. **b,** Valence band XPS spectra of $O_D$-IGZO and $O_R$-IGZO. **c,** Valence band UPS spectra of $O_D$-IGZO and $O_R$-IGZO. **d, e,** Equilibrium energy band diagrams of $O_D$-IGZO and $O_R$-IGZO before (**d**) and after (**e**) contact. **f,** Schematic of the space charge region and built-in electric field ($E_{in}$) in the $O_D$-IGZO/$O_R$-IGZO interfacial region.

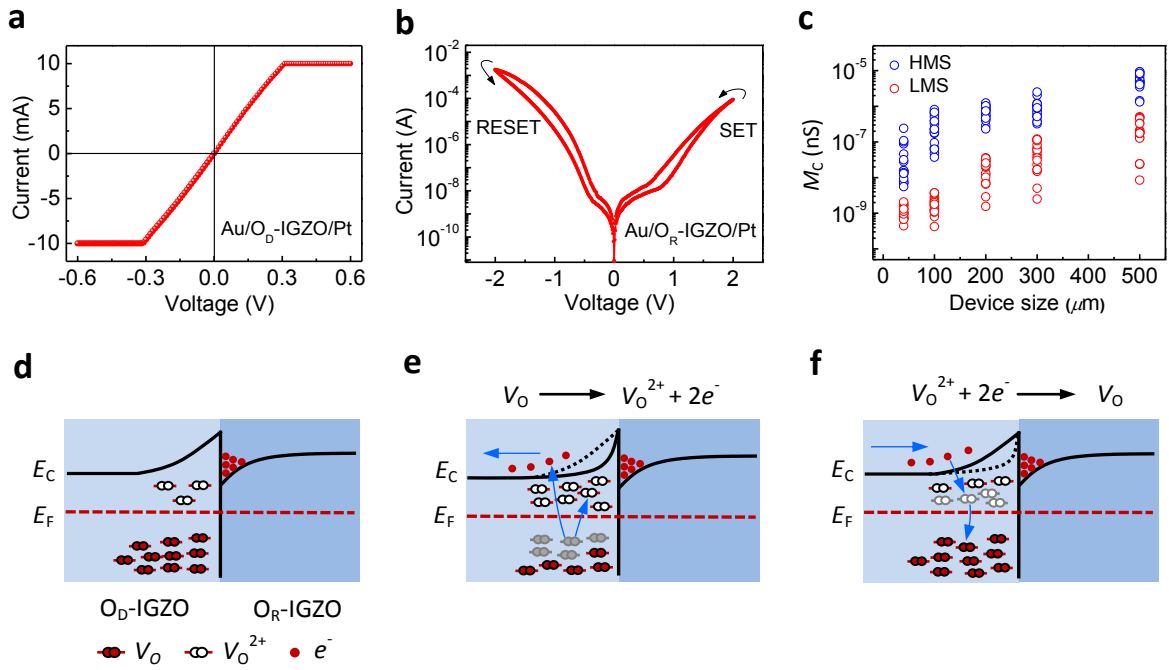

**Extended Data Fig. 4 | Memristive switching mechanism of Au/O$_D$-IGZO/O$_R$-IGZO/Pt. a, b,** *I–V* curves of Au/O$_D$-IGZO/Pt (**a**) and Au/O$_R$-IGZO/Pt (**b**) devices measured in the dark. **c,** Dependence of the memconductance of the Au/O$_D$-IGZO/O$_R$-IGZO memristor on device size. For each size, 20 memconductance values of the HMSs and LMSs measured from 20 randomly selected devices from two IGZO fabrication batches are plotted. The memconductance values were obtained at −0.2 V. The SET and RESET voltages used during the *I–V* curve tests were 2 and −2 V, respectively. **d–f,** Equilibrium energy band diagrams of the O$_D$-IGZO/O$_R$-IGZO interface in the initial LMS (**d**), after the electrical SET operation (**e**) and after the electrical RESET operation (**f**). The black dashed lines indicate the positions of $E_C$ before the SET and RESET operations. The $V_O$ ionization (blue arrows) and $V_O{}^{2+}$ neutralization (blue arrows) reactions are also schematically illustrated. Note that these reactions actually occur under nonequilibrium conditions.

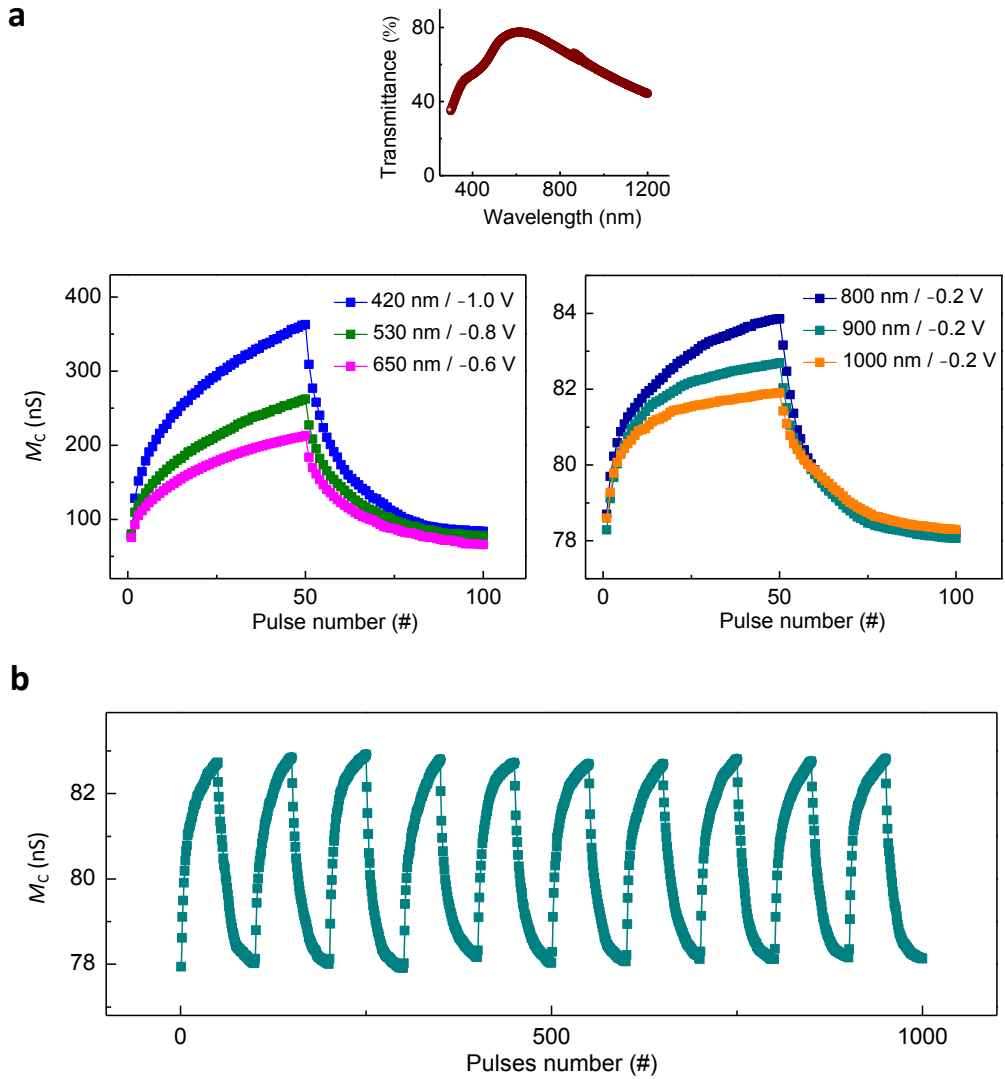

**Extended Data Fig. 5 | Au/O$_D$-IGZO/O$_R$-IGZO/Pt memristor controlled through a combination of optical and electrical stimuli. a,** Reversible modulation of the memconductance by means of 50 light pulses (optical SET) and 50 voltage pulses (electrical RESET). The inset shows the optical transmittance spectrum of Au/O$_D$-IGZO/O$_R$-IGZO/quartz. **b,** Ten successive memconductance increase–decrease cycles. The wavelength of the light pulses used for SET operations was 900 nm, and the amplitude of the electrical pulses used for RESET operations was 0.2 V. In **a** and **b**, the light pulse duration was 1 s, the light pulse interval was 1 s, the light power density was 20 $\mu$W/cm$^2$, the electrical pulse duration was 25 ms, the electrical pulse interval was 250 ms, and the memconductance values were obtained at 10 mV.

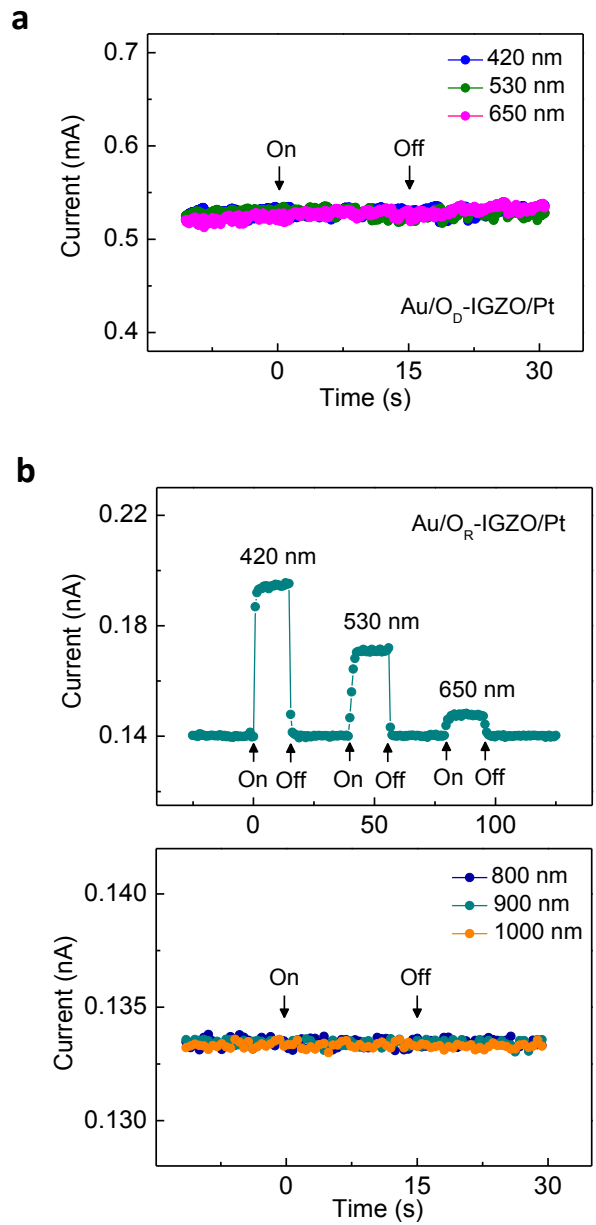

**Extended Data Fig. 6 | Optoelectronic behavior of Au/O$_D$-IGZO/Pt (a) and Au/O$_R$-IGZO/Pt (b) devices.** In **a** and **b**, the light duration was 15 s, the power density was 20 $\mu$W/cm$^2$, and the current values were measured at 10 mV.

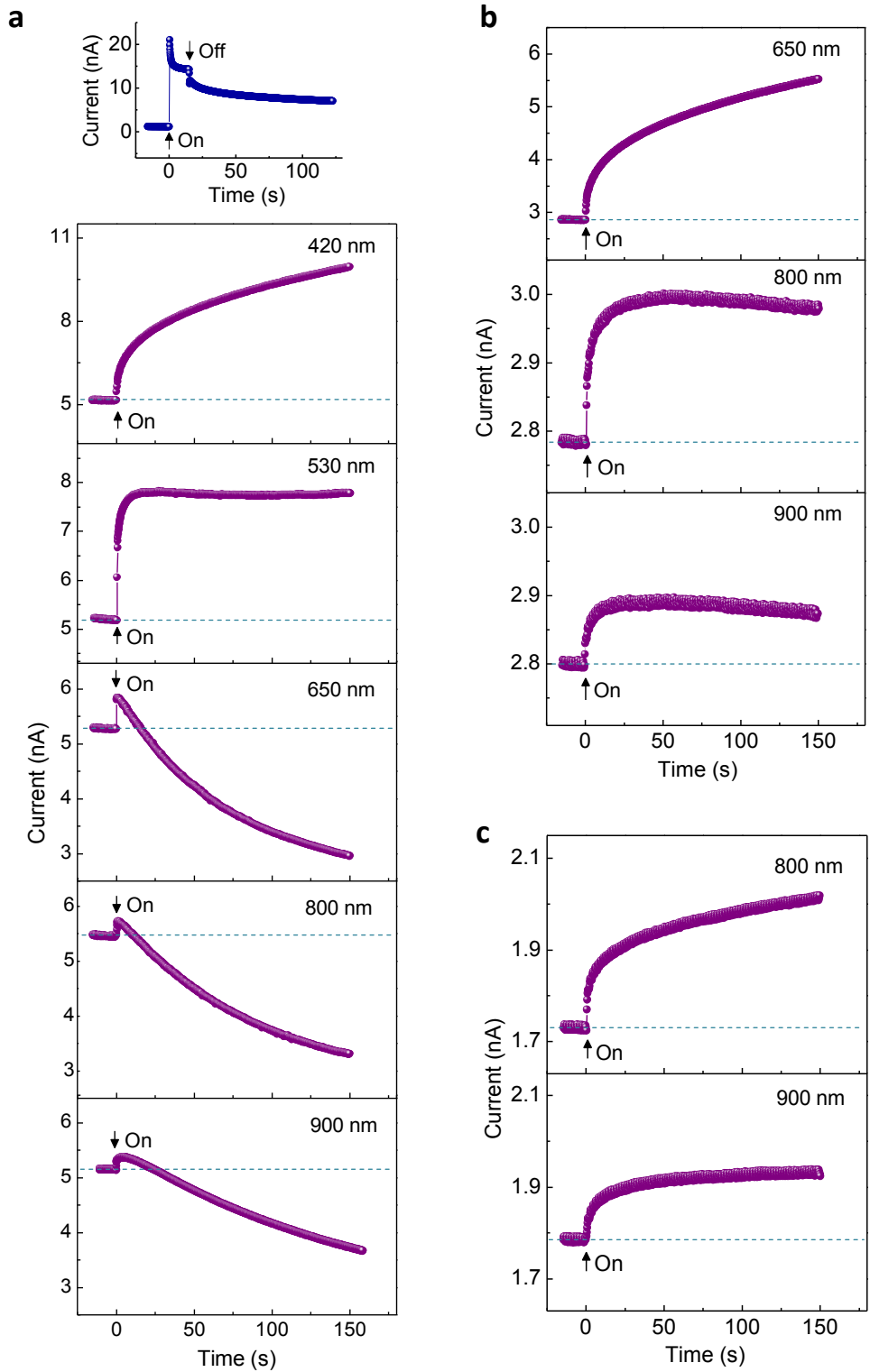

**Extended Data Fig. 7 | Photocurrent responses of a Au/O$_D$-IGZO/O$_R$-IGZO/Pt memristor to light of various wavelengths. a,** The device was first set to an HMS by ultraviolet light irradiation ($\lambda$ = 530 nm, $D$ = 15 s and $P$ = 20 $\mu$W/cm$^2$), and was then irradiated with light of longer wavelengths 10 minutes after the initial ultraviolet light exposure. The inset shows the initial SET behavior. **b,** The device was first set to an HMS by means of green light irradiation ($\lambda$ = 530 nm, $D$ = 15 s and $P$ = 20 $\mu$W/cm$^2$) and was then irradiated with red or near-infrared light ($P$ = 20 $\mu$W/cm$^2$) 10 minutes after the initial green light exposure. **c,** The device was first set to an HMS by means of red light irradiation ($\lambda$ = 650 nm, $D$ = 15 s and $P$ = 20 $\mu$W/cm$^2$) and then irradiated with near-infrared light ($P$ = 20 $\mu$W/cm$^2$) 10 minutes after the initial red light exposure. In **a**, **b** and **c**, the horizontal dashed lines indicate the initial HMS, and the current values were measured at 10 mV.

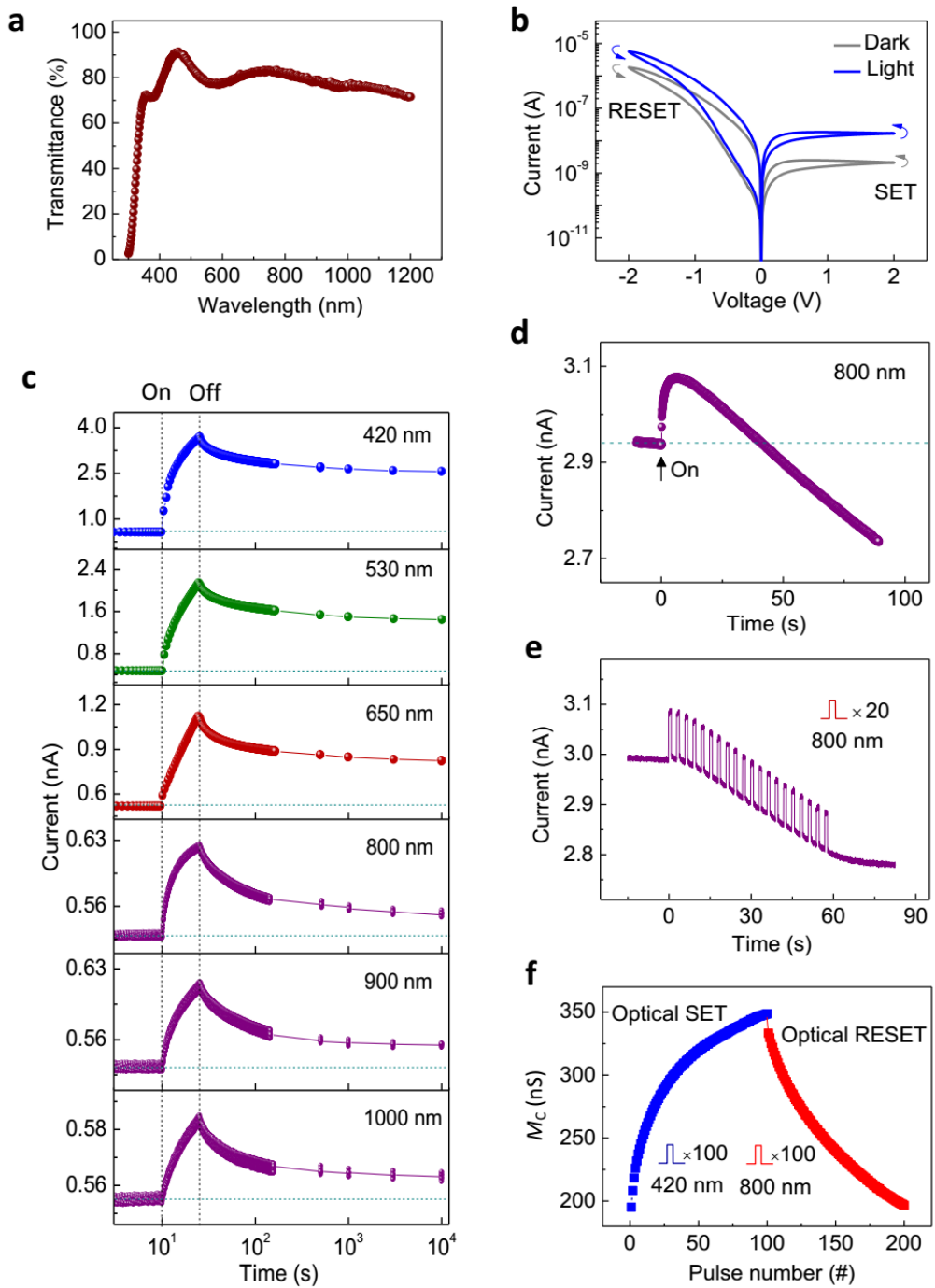

**Extended Data Fig. 8 | Optical, electrical and optoelectronic behavior of an ITO/O$_D$-IGZO/O$_R$-IGZO/Pt memristor. a,** Optical transmittance spectrum of ITO/O$_D$-IGZO/O$_R$-IGZO/quartz. **b,** $I$–$V$ characteristics before and after blue light irradiation ($\lambda = 420$ nm, $D = 15$ s and $P = 20$ $\mu$W/cm$^2$). **c,** Optical SET behavior upon irradiation with light of various wavelengths ($D = 15$ s and $P = 20$ $\mu$W/cm$^2$). The horizontal dashed lines indicate the initial LMS. The vertical dashed lines indicate the times at which the light was switched on and off. **d,** Optical RESET behavior upon continuous near-infrared light irradiation ($P = 20$ $\mu$W/cm$^2$). The device was first set to an HMS by means of blue light irradiation ($D = 15$ s and $P = 20$ $\mu$W/cm$^2$) and was then irradiated with near-infrared light 10 minutes after the initial blue light exposure. The horizontal dashed lines indicate the initial HMS. **e,** Optical RESET behavior upon near-infrared light pulses ($D = 1$ s, $I = 2$ s and $P = 20$ $\mu$W/cm$^2$). Before applying the pulses, the device underwent the same SET operation as in **d**. **f,** Reversible tuning of the memconductance by means of 100 blue light pulses ($D = 1$ s, $I = 1$ s and $P = 20$ $\mu$W/cm$^2$) and 100 near-infrared light pulses ($D = 1$ s, $I = 1$ s and $P = 20$ $\mu$W/cm$^2$). In **c**–**f**, the current and memconductance values were measured at 10 mV.

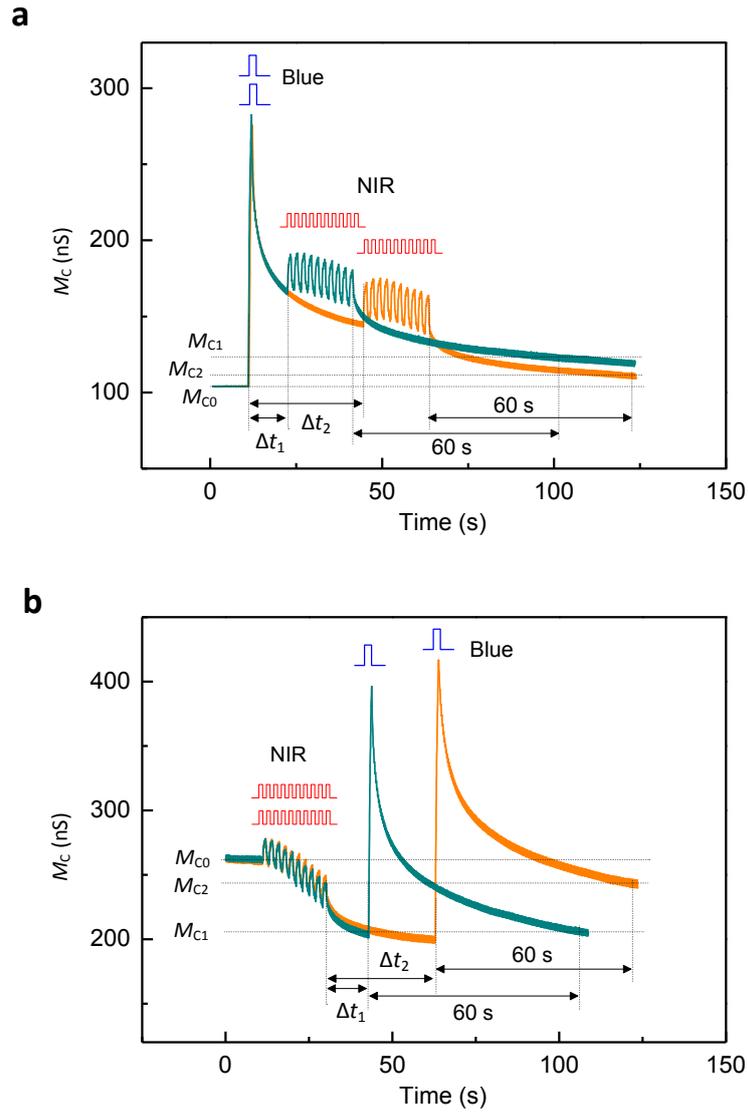

**Extended Data Fig. 9 | Mechanism of STDP emulation in a Au/O$_D$-IGZO/O$_R$-IGZO/Pt memristor. a,** When $\Delta t_1 > 0$, $\Delta t_2 > 0$ and $\Delta t_1 < \Delta t_2$, synaptic potentiation is realized ($\Delta W_1 > 0$ and $\Delta W_2 > 0$), and the degree of potentiation for $\Delta t_1$ is higher than that for $\Delta t_2$ ($\Delta W_1 > \Delta W_2$). **b,** When $\Delta t_1 < 0$, $\Delta t_2 < 0$ and $|\Delta t_1| < |\Delta t_2|$, synaptic depression is achieved ($\Delta W_1 < 0$ and $\Delta W_2 < 0$), and the degree of depression for $\Delta t_1$ is higher than that for $\Delta t_2$ ($|\Delta W_1| > |\Delta W_2|$). The insets illustrate the light pulse schemes. In **a** and **b**, the memconductance values were obtained at 10 mV, and NIR represents near-infrared light.